\documentstyle[amscd,amssymb,verbatim,12pt]{amsart}
\def\dspace{\baselineskip=0.3 in}
\begin{document}
\dspace
\title[Early and Late Transient....]{EARLY AND LATE TRANSIENT COSMIC
  ACCELERATION DUE TO CURVATURE INSPIRED DARK ENERGY }

\author{\bf S.K.Srivastava}
{ }
\maketitle
\centerline{ Department of Mathematics,}
 \centerline{ North Eastrn Hill University,}
 \centerline{  Shillong-793022, India}
\centerline{ srivastava@@.nehu.ac.in; sushil@@iucaa.ernet.in }

\vspace{1cm}

It is obtained that dark energy emerges
from the higher-derivative gravity with non-linear terms proportional to $R^2$ and
$R^3$ with $R$ being the Ricci scalar curvature. Interestingly,
it is found that the universe begins with acceleration, which continues for
a short period. Later on, it decelerates driven by radiation and
subsequently 
by matter. Two components of dark energy are found here. In the early
universe, dark energy mimics quintessence giving high 
value of initial density $\sim 10^{75} {\rm GeV}^4$ at Planck scale. But, in
the late universe, dark energy behaves like phantom giving current value of its
density and causing late  acceleration for some time. Thereafter, deceleration
driven by matter resumes. Thus, two components of dark energy are obtained
here (i) quintessence-like in the early universe and (ii) phantom-like in the
late universe.  Moreover, it is
interesting to see acceleration being a
transient phenomonon in early as well as late universe.

\vspace{2cm} 
 
Modifying earlier results \cite{sp}, observations of 16 Type SNe Ia show a
conclusive evidence for deceleration preceding acceleration in the late
universe \cite{ag}. Cosmic acceleration is fuelled by a source of energy
violating strong (SEC) or weak energy condition (WEC), called dark energy
(DE).  Instead of the simplest $\Lambda$-model ($\Lambda$ being the
cosmological constant) having fine-tuning problem, dynamical models, given by
different scalars (quintessence, tachyon and phantom) were envisaged in the
past few years to explain late cosmic acceleration and fall of DE density
(DED) from a very high to current low value \cite{vs}. Due to negative pressure
and being better fit with WMAP, generalized Chaplygin gas (GCG) have been a
good candidate for DE \cite{ob}. In \cite{ms}, another attempt is made adding
string gravity corrections with dilaton or modulous fields to gravitational
action. Recently, some gravitational models are proposed, where non-linear
term of the Ricci scalar $R$ is taken as DE and its various
cosmological consequences are discussed \cite{snsd}.

In these models, non-gravitational
or gravitational source of DE is taken {\em a priori}. But, it is more
interesting to have model, where DE emerges spontaneously from the
gravitational sector.

In what follows,, non-linear term of curvature
is {\em not} taken as DE lagrangian {\em a priori}, which is  unlike the approach in \cite{snsd} . Rather, in the modified
Friedmann equations (MFE), obtained here, DE density terms emerge due to
presence of linear as well as non-linear terms of curvature scalar in the
action. This is the approach adapted in \cite{sksde}. 

In modern cosmology, universe is supposed to have {\em acceleration} in
the early universe followed by {\em deceleration} driven by radiation and subsequently by matter as well as {\em late acceleration} driven
by DE, preferably phantom-like. Here, a cosmological model is proposed
yielding this cosmic scenario. In addition to DE terms (mentioned above), here, it is found that
dark matter (DM) also emerges from the gravitational sector. For the radiation
term another scalar $\phi$ is introduced. It is found that this term
emerges as a combined effect of  $\phi$ and the scalar curvature $R$.

Here, it is found that DE is induced by curvature
\cite{sksde} causing early and
late cosmic acceleration, which are {\em transient phenomena}. Interestingly,
DE, obtained here, has two components (i) quintessence-like giving high value
of 
DED in the early universe and (ii)phantom-like in the late universe giving
current value of DED. This model is interesting, in the sense, that it
explains different phases of the cosmic dynamics i.e. (i) inflation in the
early universe, (ii) deceleration after exit from inflationary phase,  (iii)
transition from deceleration to acceleration in the late universe and (iv)
another transition from acceleration to deceleration in the future
universe. Natural units $({\hbar} = c = 1)$ are used here with GeV as a
fundamebtal unit and $1 {\rm GeV}^{-1} = 6.54\times 10^{-25} {\rm sec}.$

As mentioned above, modifications of gravitational action with higher powers
of scalar curvature $R$ is not new. So it is natural to probe cosmological
consequences on taking $R^3$ in the gravitational action. Earlier, Amendola \cite{am}
also had taken gravitational action having lagrangian $R + R^3$. Here,
investigations begin from the Planck scale.
 
The action is taken as
$$ S = \int {d^4x} \sqrt{-g} \Big[\frac{R}{16 \pi G} + \alpha R^2 - \beta R^3 
+ \{\frac{1}{2} g^{\mu\nu} \triangledown_{\mu}
\phi(x) \triangledown_{\nu}\phi(x) - V(\phi(x)) \} \Big],  \eqno(1)$$
where $G = M_P^{-2} (M_P = 10^{19} {\rm GeV}$ is the Planck mass), $\alpha$ is
a dimensionless coupling constant, $\beta$ is a constant having dimension
(mass)$^{-2}$ (as $R$ has mass dimension 2)and $\triangledown_{\mu}$ stands for covariant derivative. Moreover, $\phi(x)$ is a scalar field with
$V(\phi)$ as potential giving radiation term in the modified Friedmann
equation obtained here.

Action (1) yields field equations
$$\frac{1}{16 \pi G} (R_{\mu\nu} - \frac{1}{2} g_{\mu\nu} R) + \alpha (2
\triangledown_{\mu} \triangledown_{\nu} R - 2 g_{\mu\nu} {\Box} R -
\frac{1}{2} g_{\mu\nu} R^2 + 2 R R_{\mu\nu} ) $$
$$ - \beta (3 \triangledown_{\mu} \triangledown_{\nu} R^2 - 3 g_{\mu\nu} {\Box}
R^2 - \frac{1}{2} g_{\mu\nu} R^3 + 3 R^2 R_{\mu\nu} ) $$
$$  + \frac{1}{2}
\Big\{\triangledown_{\mu}\phi(x) \triangledown_{\nu}\phi(x) -
g_{\mu\nu}\Big(\frac{1}{2} \triangledown^{\sigma}
\phi(x) \triangledown_{\sigma}\phi(x) - V(\phi(x))\Big) \Big\}   = 0
\eqno(2a)$$ 
using the condition $\delta S/\delta g^{\mu\nu} = 0$. Moreover, invariance of
$S$ under transformation $\phi \to \phi + \delta \phi$ yields
$$ {\Box} \phi + V^{\prime}(\phi) = 0 , \eqno(2b)$$
where $V^{\prime}(\phi_i) = dV(\phi)/d\phi$ and the operator
$$ {\Box} = \frac{1}{\sqrt{-g}} \frac{\partial}{\partial x^{\mu}}
\Big(\sqrt{-g} g^{\mu\nu} \frac{\partial}{\partial x^{\nu}} \Big). \eqno(2c)$$

Taking trace of (2a), it is obtained that
$$ - \frac{R}{16 \pi G} - 6 (\alpha - 3 \beta R){\Box} R + 18 \beta
\triangledown^{\mu}R \triangledown_{\mu}R - \beta R^3$$
$$ + \frac{1}{2} [ - \triangledown^{\sigma}
\phi(x) \triangledown_{\sigma}\phi(x) + 4 V(\phi(x))] = 0 \eqno(3)$$

In (3), $(\alpha - 3 \beta R)$ emerges as a coefficient of ${\Box} R$ due to
presence of terms $\alpha R^2$ and $\beta R^3$ in the action (1). If $\alpha =
0$, effect of $R^2$ vanishes and  effect of $R^3$ is switched
off for $\beta = 0$. So, an {\em effective} scalar curvature ${\tilde R}$ is defined as
$$ \gamma {\tilde R} = \alpha - 3 \beta R , \eqno(4a)$$
where  $\gamma$ is a constant having dimension (mass)$^{-2}$ being
used for dimensional correction. For $t > t_P$ ($t_P$ being the Planck time),
scalar curvature at Planck scale $R_P > R$. Here $\alpha$ and $\beta$ are
taken so that $(\alpha - 3 \beta R_P) > 0$, giving
$$ (\alpha - 3 \beta R) > (\alpha - 3 \beta R_P) > 0. \eqno(4b)$$

Connecting (3) and (4a), it is obtained that
$$ - {\Box}{\tilde R} - \frac{1}{\tilde R}\triangledown^{\mu}{\tilde R}
\triangledown_{\mu} {\tilde R} = \frac{1}{6 \gamma} \Big[\frac{1}{16 \pi G} +
\frac{\alpha^2}{3\beta} \Big] - \frac{1}{6 \gamma} \Big[\frac{1}{16 \pi G} +
\frac{\alpha^2}{9\beta} \Big]\frac{1}{\gamma{\tilde R}}$$
$$ + \frac{\tilde R}{54 \beta} [\gamma {\tilde R} - 3 \alpha ] + \frac{\beta}{4
  \gamma^2 {\tilde R}}[ - \triangledown^{\sigma}
\phi(x) \triangledown_{\sigma}\phi(x) + 4 V(\phi(x))]  \eqno(5)$$

Energy-momentum tensor components for $\phi$ are
$$ T^{\mu\nu}_{\phi} = \triangledown^{\mu}\phi \triangledown^{\nu}\phi -
g^{\mu\nu}\Big(\frac{1}{2} \triangledown^{\sigma}
\phi \triangledown_{\sigma}\phi - V(\phi)\Big) \eqno(6a)$$
yielding energy density for $\phi$ as
$$ \rho_{\phi} = \frac{1}{2}\triangledown^{\sigma}
\phi \triangledown_{\sigma}\phi  + V(\phi)  \eqno(6b)$$
and pressure for $\phi$ as
$$ p_{\phi} = \frac{1}{2} \triangledown^{\sigma}
\phi \triangledown_{\sigma}\phi - V(\phi).  \eqno(6c)$$

 Experimental evidences \cite{ad} support spatially homogeneous
flat model of the universe 
$$dS^2 = dt^2 - a^2(t) [dx^2 + dy^2 + dz^2] \eqno(7)$$
with $a(t)$ being the scale-factor.

In this space-time, (2b), (5),(6b) and (6c) are obtained as
$$ {\ddot \phi} + 3 \frac{\dot a}{a} {\dot \phi} + V^{\prime}(\phi) = o, \eqno(8a)$$
$$- {\ddot {\tilde R}} - 3 \frac{\dot a}{a} {\dot {\tilde R}} -\frac{{\dot
{\tilde R}}^2}{{\tilde R}} =  \frac{1}{6 \gamma} \Big[\frac{1}{16 \pi G} +
\frac{\alpha^2}{3\beta} \Big] - \frac{1}{6\gamma} \Big[\frac{1}{16 \pi G} +
\frac{\alpha^2}{9\beta} \Big]\frac{1}{\gamma{\tilde R}}$$
$$ + \frac{\tilde R}{54 \beta} [\gamma {\tilde R} - 3\alpha ] + \frac{\beta}{4
  \gamma^2 {\tilde R}}\Big[ - {\dot \phi}^2 + 4 V(\phi)\Big], \eqno(8b)$$
$$ \rho_{\phi} = \frac{1}{2}{\dot \phi}^2 + V(\phi)  \eqno(8c)$$
and 
$$ p_{\phi} = \frac{1}{2}{\dot \phi}^2 - V(\phi)  \eqno(8d)$$
as $\phi(x, t) = \phi(t)$ due to spatial homogeneity.

For $a(t)$, being the power-law function of cosmic time, ${\tilde R} \sim a^{-n}$. For
example, ${\tilde R} \sim a^{-4}$ for radiation model and ${\tilde R} \sim a^{-3}$ for
matter-dominated model. So, there is no harm in taking
$$  {\tilde R} = \frac{A}{a^n} , \eqno(9)$$
where $n > 0$ is a real number and $A$ is a constant with mass dimension 2.

(8a) integrates to
$$ \frac{d}{dt}\Big[\frac{1}{2}{\dot \phi}^2 + V(\phi) \Big] + 3
\frac{\dot a}{a} {\dot \phi}^2 = 0. \eqno(10a)$$

For a cosmic fluid, pressure $p$ is proportional to its energy density
$\rho$. So, for fluid of scalar $\phi$, the equation of state is taken as
$$ p_{\phi} = k \rho_{\phi},$$
where $k$ is a  proportionality constant.
Also, from (8c) and (8d), we obtain
$$ V(\phi) = \frac{1}{2} \Big(\frac{1 - k}{1 + k} \Big) {\dot
  \phi}^2. \eqno(10b)$$

In what follows, we find very intersting results, on taking $k$ suitably, as
$$  k = \frac{1 + 2n}{3}. \eqno(10c)$$
Using these values of $k$ in (10a), we obtain 

$$ V(\phi) = \frac{(1 -  n)}{2 (2 + n)} {\dot \phi}^2, \eqno(10d)$$

Using (10c) in (10a)  and integrating, we obtain 
$${\dot \phi}^2 = \frac{B}{a^{(2n + 4)}} . \eqno(11)$$
Here $B$ is an integration constant.

From (10d) and (11)

$$ - {\dot \phi}^2 + 4 V(\phi) = - \frac{3 n}{(2 +  n)} \frac{B}{a^{(2n + 4)}} . \eqno(12)$$

Using (9) and (12) in (8b) ,it is obtained that

$$\frac{d}{dt} \Big(\frac{\dot a}{a}\Big) + (3 - 2 n) \Big(\frac{\dot a}{a}
\Big)^2 = \frac{ C a^n}{n A } \Big[ 1 - \frac{D a^n}{C} \Big] +
\frac{\gamma A}{54 n \beta} \Big[ \frac{1}{a^n} - \frac{3 \alpha}{\gamma A}
\Big]$$
$$ -
\frac{3 \beta B}{4 (2 + n) \gamma^3 A^2 a^4},  \eqno(13a)$$
where
$$ C = \frac{1}{6 \gamma} \Big[\frac{1}{16 \pi G} + \frac{\alpha^2}{3 \beta}
\Big]    \eqno(13b)$$
and
$$ D = \frac{1}{6 \gamma} \Big[\frac{1}{16 \pi G} + \frac{\alpha^2}{9 \beta}
\Big].    \eqno(13c)$$

(13a) can be re-written as
$${\ddot a} + (2 - 2 n) \frac{{\dot a}^2}{a}
 = \frac{ C a^{n + 1}}{n A } \Big[ 1 - \frac{D a^n}{C} \Big] +
\frac{\gamma A a}{54 n \beta} \Big[ \frac{1}{a^n} - \frac{3 \alpha}{\gamma A}
 \Big]$$
$$ -
\frac{3 \beta B}{4 (2 + n) \gamma^3 A^2 a^3},  \eqno(14)$$
which integrates to
\begin{eqnarray*}
 \Big(\frac{\dot a}{a} \Big)^2 &=& \frac{E}{a^{6 - 4n}} + \frac{2 C}{n A
  } \Big[ \frac{a^n}{(6 - 3n)} - \frac{D a^n}{C (6 - 2n)} \Big]\\&&
 + \frac{\gamma A }{27 n \beta} \Big[ \frac{1}{(6 - 5n)a^n} - \frac{3
  \alpha}{\gamma A (6 - 4n)} \Big]\\&&
 - \frac{3 \beta B}{2 (2 + n)(2 - 4n) \gamma^3 A^2 a^4},
\end{eqnarray*}
\vspace{-1.8cm}
\begin{flushright}
 (15a)
\end{flushright}
\vspace{0.2cm}
where $E$ is an integration constant having dimension (mass)$^2$.

(15a) is the modified Friedmann equation (MFE) giving cosmic dynamics. In this
equation, the term proportional to $a^{-4}$ is the combined effect of
curvature and scalar field $\phi$. Other terms, on r.h.s. of this equation,
are imprints of curvature. The first term of these, proportional to $a^{- (6
  -4n)}$ emerge spontaneously. It is interesting to see that this term (the
first term on r.h.s. of (15a)) corresponds to matter density if $n = 3/4$
i.e. for this value of $n$, it reduces to $E a^{-3}$. As it is spontaneously
created from the gravitational sector, we recognize it as as dark matter
density.

For $n = 3/4$, (15a) looks like 
\begin{eqnarray*}
 \Big(\frac{\dot a}{a} \Big)^2 &=& \frac{E}{a^3} + \frac{32 C a^{3/4}}{45 A
  } \Big[ 1 - \frac{5 D a^{3/4}}{6 C} \Big] \\&&+
\frac{ 16 \gamma A }{729 \beta} \Big[ \frac{1}{a^{3/4}} - \frac{ \alpha}{\gamma A} \Big] +
\frac{6 \beta B}{11 \gamma^3 A^2 a^4}.  
\end{eqnarray*}
\vspace{-1.8cm}
\begin{flushright}
 (15b)
\end{flushright}
\vspace{0.2cm}

On r.h.s. of (15b), terms proportional to $a^{3/4} (a^{- 3/4})$ are recognized
as dark energy density being created by curvature. Moreover, the term
proportional to $a^{-4}$ corresponds to radiation as usual.

(15b) is re-written as
 $$ \Big(\frac{\dot a}{a} \Big)^2 =  \frac{32 C a^{3/4}}{45 A
  } \Big[ 1 - \frac{5 D a^{3/4}}{6 C} \Big] +
\frac{ 16 \gamma A }{729 \beta} \Big[ \frac{1}{a^{3/4}} - \frac{ \alpha}{\gamma A}
  \Big] + \frac{8 \pi G}{3} (\rho_{\rm rd} + \rho_{\rm dm})   \eqno(15c)$$
with
$$ \rho_{\rm rd} = \frac{9 \beta B}{44 \gamma^3 A^2 a^4}  \eqno(15d)$$
and
$$ \rho_{\rm dm} =  \frac{3 E}{8 \pi G a^3}. \eqno(15e)$$

Using current values of $\rho_{\rm rd}$ and
$\rho_{\rm dm}$ as $5 \times 10^{-5} \rho_{\rm cr}$ and $0.23 \rho_{\rm cr}$ 
respectively \cite{abl}, (15d) and (15e) yield 
$$\rho_{\rm dm} = 0.23 \rho_{\rm cr} \Big(\frac{a_0}{a} \Big)^3
\eqno(16a)$$
and
$$\rho_{\rm rd} = 5 \times 10^{-5} \rho_{\rm cr} \Big(\frac{a_0}{a} \Big)^4,
\eqno(16b)$$
where $a_0 = a(t_0), {3 E}/{8 \pi G } =  0.23 \rho_{\rm cr} a_0^3, {9
  \beta B}/{44 \gamma^3 A^2 } = 5 \times 10^{-5} \rho_{\rm cr}a_0^4  $  and  
$$ \rho_{\rm cr} = \frac{3 H_0^2}{8 \pi G}  $$
with $H_0$ being the current Hubble's rate.

(16a) and (16b) look like
$$\rho_{\rm dm} = \frac{0.23 \rho_{\rm cr}}{a^3} \eqno(16c)$$
and
$$\rho_{\rm rd} = \frac{5 \times 10^{-5} \rho_{\rm cr}}{a^4} \eqno(16d)$$
on normalizing $a_0$ as
$$ a_0 = 1. \eqno(16e)$$

(16c) and (16d) yield that $\rho_{\rm rd} < \rho_{\rm m}$ at red-shift
$$ Z = \frac{1}{a} - 1 < 5399. $$
WMAP \cite{abl} gives decoupling of matter from radiation at red-shift
$$ Z_d = \frac{1}{a_d} - 1 = 1089. \eqno(17)$$
It shows that dark matter dominates over radiation when $a > a_d$.

\bigskip
\noindent Case 1. \underline{When  $ a < 2.17\times 10^{-4} $}
\smallskip

This is the case of {\em early} universe, where (15c) reduces to

$$ \Big(\frac{\dot a}{a} \Big)^2 = \frac{8 \pi G}{3} \Big[ \Big\{\rho^{\rm
  qu}_{\rm  de} - {2 \lambda^{\rm qu}} \Big\} +
  \rho_{\rm rd} \Big]  \eqno(18)$$
as term $\sim a^{- 3/4}$ dominates over term $\sim a^{3/4}$. In (18),
$$\rho^{\rm qu}_{\rm  de} = \frac{16 \gamma A}{729 \pi G \beta} a^{-3/4}  \eqno(19a)$$
and
$$\lambda^{\rm qu} = \frac{ \alpha }{243 \pi G \beta} \eqno(19b)$$
being the {\em cosmic tension} in the early universe, which is caused by
scalar curvature.

The conservation equation for DE is
$${\dot \rho_{\rm de}} + 3 \frac{\dot a}{a} (1 + {\rm w}_{\rm de}) \rho_{\rm
  de} = 0, \eqno(20 a)$$
where equation of state parameter (EOSP) ${\rm w}_{\rm de} = p_{\rm
  de}/\rho_{\rm de}$ with $p_{\rm de}$ being DE fluid pressure  . (19a) and (20a)
  yield
$$  {\rm w}^{\rm qu}_{\rm de} = - \frac{3}{4}  \eqno(20 b)$$
 It shows that DE component,
dominating the early universe, mimics quintessence.

At the Planck scale, $\rho_{\rm de} =
      ({3/}{8 \pi G}) M_P^2 ,$ giving 
$$ \rho^{\rm qu}_{\rm de} = \frac{3}{8 \pi G} M_P^2 \Big(\frac{a_P}{a}
\Big)^{3/4}  \eqno(21)$$
with $a_P = a(t_P)$ and $t_P = M_P^{-1}$ being the Planck time.

Using (16d), (19a) and (19a) in (18), it is obtained that
$$ \Big(\frac{\dot a}{a} \Big)^2 = \frac{16 \gamma A}{729 \beta a^{3/4}}
\Big[1 - \frac{\alpha}{\gamma A} a^{3/4} \Big] + \frac{5 \times 10^{-5}
  H_0^2}{a^4} . \eqno(22a)$$ 
This equation shows that at 
$$a = a_{\rm qe} = \Big(\frac{\gamma A}{\alpha})\Big)^{4/3}, \eqno(22b)$$
terms within the bracket on r.h.s. vanishe and, for $a >  a_{\rm qe},$
expansion of the universe is driven by radiation.

In the very early universe, first term on r.h.s of (22a) dominates over the
third term, so (22a) is obtained as
$$ \Big(\frac{\dot a}{a} \Big)^2 = \frac{16 \gamma A}{729 \beta a^{3/4}}
\Big[1 - \frac{\alpha}{\gamma A} a^{3/4} \Big], \eqno(22c)$$ 
which integrates to
$$ a^{3/8} = a_P^{3/8} \Big(\sqrt{\frac{\alpha}{\gamma A}}\Big)^{-1} sin \Big[\sqrt{2 \pi G\lambda^{\rm qu}} (t - t_P) +
sin^{-1} \Big(\sqrt{\frac{\alpha a_P^{3/4}}{\gamma A} }\Big) \Big]. $$
As, in the early universe, $(t - t_P)$ is very small and $\sqrt{{\alpha a_P^{3/4}}/{\gamma A} }$ is also small, so this solution is apprximated as
$$ a(t) \simeq  a_P \Big[\sqrt{(2 \pi G \gamma A/\alpha) \lambda^{\rm qu}}(t - t_P) + 1 \Big]^{8/3} \eqno(22d)$$
using $ sin x \simeq x$ for small $x$. 

Also, at Planck scale, $ \rho^{\rm qu}_{\rm de(P)} > 2 \lambda^{\rm qu}$. So,
from eqs.(19a), (19b),(22b) and inequality (4b)
$$ 3 \beta R_P < \alpha < \frac{8 \gamma A}{3 a_P^{3/4}}.$$

Connecting (22b) and (22d)
$$ a(t) \simeq  a_P \Big[\sqrt{2 \pi G \lambda^{\rm qu}(a_{\rm qe}/ a_P)^{3/4}}(t - t_P) + 1 \Big]^{8/3}. \eqno(22e)$$

(22e) gives power-law inflation (acceleration)
in the early universe. This scenario continues till $\rho^{\rm
  qu}_{\rm  de} = 2 \lambda^{\rm qu}$ upto the time  
$$t_{\rm qe} = t_P + 2 t_P \sqrt{\frac{2}{3}}
\Big[\Big(\frac{a_{\rm qe}}{a_P} \Big)^{3/8} - 1 \Big] \eqno(23a)$$
as
$$\frac{M_P^2}{16 \pi G \lambda^{\rm qu}} = \Big(\frac{a_{\rm qe}}{a_P}
\Big)^{3/4} .\eqno(23b)$$
By the time $t_{\rm qe}$ the scale factor grows to $a_{\rm qe}$.
If the early universe inflates sufficiently during this period and goes over
to 65 e-foldings,
$$ \frac{a_{\rm qe}}{a_P} \simeq 10^{28}   \eqno(23c)$$
and
$$ t_{\rm qe} - t_P \simeq 2 t_P \sqrt{\frac{2}{3}}
\Big(\frac{a_{\rm qe}}{a_P} \Big)^{3/8} = 5.16\times 10^{10} t_P = 3.37\times
10^{-27} {\rm sec.} . \eqno(23d)$$

Connecting (23b) and (23c), it is obtained that
$$ \lambda^{\rm qu} =  \frac{3}{16 \pi G} M_P^2 \Big(\frac{a_P}{a_{\rm qe}}
\Big)^{3/4} = 5.97 \times 10^{53} {\rm GeV}^4  \eqno(23e)$$

For $t_d > t > t_{\rm qe}$, using $\rho^{\rm qu}_{\rm  de} = 2 \lambda^{\rm qu}$ (22a) is obtained as
$$\Big(\frac{\dot a}{a} \Big)^2 = 5 \times 10^{-5} H_0^2 \Big(\frac{1}{a}
\Big)^4, $$
which integrates to
$$ a(t) = a_{\rm qe} \Big[1 + 0.14142 H_0 \Big(\frac{1}{a_{\rm qe}} \Big)^2 (t -
t_{\rm qe}) \Big]^{1/2} \eqno(24)$$
giving deceleration as ${\ddot a } < 0.$

\newpage

\noindent Case 2. \underline{When $ a > 2.17\times 10^{-4} $} 

\smallskip

This is the case of {\em late} universe, where (15a) reduces to
$$\Big(\frac{\dot a}{a} \Big)^2 = \frac{8 \pi G}{3} \Big[\rho^{\rm ph}_{\rm de}
\Big\{1 -  \frac{\rho^{\rm ph}_{\rm de}}{2 \lambda^{\rm ph}}  \Big\} +
\rho_{\rm  dm} \Big] \eqno(25)$$
as term $\sim a^{3/4}$ dominates over term $\sim a^{- 3/4}$. Here 

$$\rho^{\rm ph}_{\rm de} = \frac{4 C}{15 \pi A G} a^{3/4}        \eqno(26)$$

and
$$ \lambda^{\rm ph} = \frac{3 C^2}{25 \pi G D A}        . \eqno(27)$$

Equation, like (25) with $\rho^2$ term, also arise in brane-gravity inspired
Friedmann equation with $\lambda$ called as brane-tension \cite{rm}. Here,
this term arises from the 4-dimensional higher-derivative gravity without using
any prescription of the brane-world. $\lambda^{\rm ph}$ is called {\em cosmic
  tension} in the late universe.

Connecting (26) and (20a)
$$  {\rm w}^{\rm ph}_{\rm de} = -5/4  ,\eqno(28)$$
which shows that DE component
dominant in the late universe mimics phantom.

Current value of $\rho_{\rm de}$ is $0.73 \rho_{\rm cr}$ \cite{abl}, so (26) reduces to
$$\rho^{\rm ph}_{\rm de} = 0.73 \rho_{\rm cr} {a}^{ 3/4}  \eqno(29)$$
with ${4 C}/{15 \pi A G} = 0.73 \rho_{\rm cr}$. 
For $ a < a_0 = 1, \rho^{\rm ph}_{\rm de} > \Big(\rho^{\rm ph}_{\rm de} \Big)^2.$
      So, for $\rho_{\rm dm} > \rho^{\rm ph}_{\rm de},$ (25) integrates to
$$ a(t) = a_d [ 1 + 25887.8 H_0 (t - t_d) ]^{2/3} \eqno(30)$$
for $(1/a_d)$, given by (17) and $\rho_{\rm dm}$ given by (16a). (30) gives ${\ddot a} < 0$ showing deceleration.

(16a) and (26) show transition from $\rho_{\rm m} > \rho^{\rm ph}_{\rm de}$ to
      $\rho_{\rm m} < \rho^{\rm ph}_{\rm de}$ at red-shift
$$ Z_* = \frac{1}{a_*} - 1 = \Big(\frac{73}{23} \Big)^{4/15} - 1 = 0.36 \eqno(31)$$
with $a_* = a(t_*)$ and $t_*$ being the transition time. Thus,
red-shift,corresponding to the transition time $t_*$, is obtained within the
range $0.33 \le Z_* \le 0.59$ given by 16 Type supernova observations \cite{ag}.

Thus, for $t \ge t_*$, using (26) and (28), (25) reduces to
$$\Big(\frac{\dot a}{a} \Big)^2 = \frac{8 \pi G}{3} \rho^{\rm ph}_{\rm de}
\Big\{1 -  \frac{\rho^{\rm ph}_{\rm de}}{2 \lambda^{\rm ph}}  \Big\},  $$
which integrates to
\begin{eqnarray*}
 a(t) &=&  \Big[ \frac{0.73 \rho_{\rm cr}} {2 \lambda^{\rm ph}} + 
\Big\{\sqrt{ 1.26 - \frac{0.73 \rho_{\rm cr}} {2 \lambda^{\rm ph}}}
\\ &&
- \frac{3}{8} H_0 \sqrt{0.73} (t - t_*)
\Big\}^2 \Big]^{- 4/3}
\end{eqnarray*}
\vspace{-1.7cm}
\begin{flushright} 
(32)
\end{flushright}

\vspace{0.3cm}
using (29) and (31).

(32) shows acceleration till $\rho^{\rm ph}_{\rm de}$ becomes equal to $2
\lambda^{\rm ph}$ as it grows with
expansion. It happens till $a(t)$ increases to $a_{\rm pe}$ satisfying
$$ {a_{\rm pe}}^{3/4} = \frac{2
  \lambda^{\rm ph}}{0.73 \rho_{\rm cr}}. \eqno(33)$$
Thus, accelerated growth of $a(t)$, given by (32), stops at time
$$ t_{\rm pe} = t_*  + \frac{8}{3 H_0 \sqrt{0.73}} \sqrt{1.26 - \frac{0.73 \rho_{\rm cr}}{2 \lambda^{\rm ph}}} \eqno(34)$$
and (25) reduces to
$$\Big(\frac{\dot a}{a} \Big)^2 = \frac{8 \pi G}{3} \rho_{\rm  dm} \eqno(35)$$
yielding 
$$ a(t) = a_{\rm pe}[ 1 + 0.72 H_0 a_{\rm pe}^{-3/2}(t - t_{\rm pe}) ]^{2/3} \eqno(36)$$

Thus, late acceleration is obtained  during the period $(t_{\rm pe} - t_*)$. For $t > t_{\rm pe}$,
     DE is not able to drive cosmic dynamics as $\rho^{\rm ph}_{\rm de} > 2
     \lambda^{\rm ph}$ leads to DED component negative in (25) and $a(t)$
     complex.  So deceleration, driven by dark matter,
     will resume. Interestingly, solution (32) does not fall into {\em finite    time future singularity} \cite{sn05}. Here, late transient cosmic acceleration is obtained
     using the modified gravity. In \cite{vs02}, this phenomenon is obtained using RS2 brane-model.
     .

Thus, it is found that DE, emerging due to curvature,  causes {\em transient} acceleration in the early and
late universe . It is {\em transient} as {\em cosmic tension}
puts a {\em brake} on accelerated expansion. Interestingly, in early and late
universe, {\em cosmic tension} also arises from the gravitational sector. Moreover, it is interesting to
see that quintessence dark energy emerges in the very early universe, where
scalar curvature is high and phantom dark energy emerges in the late universe, where
scalar curvature is low. Thus, according to the present model, low scalar curvature
produces phantom dark energy and high curvature produces  quintessence dark
energy.

\bigskip

\centerline{\bf Acknowledgement}

Author is grateful to Prof. S.D.Odinstov for his useful remarks and
suggestions, which improved this paper.

\end{document}